\documentclass[12pt]{article}
\usepackage{hyperref}
\hypersetup{colorlinks,urlcolor=blue,citecolor=blue,linkcolor=blue,filecolor=black}
\usepackage{graphicx}
\usepackage{amsmath}
\usepackage{bm}
\usepackage{amssymb}
\usepackage{footmisc}

\def\HS{S}

\newcommand{\w}{\omega}
\newcommand{\y}{y}
\newcommand{\n}{n}
\newcommand{\signN}{(-1)^r}

\newcommand{\signj}{(-1)^j}
\newcommand{\z}[1]{\zeta_{#1}}

\newcommand{\hh}[2]{\mathrm{h}_{#1#2}}
\newcommand{\hhh}[3]{\mathrm{h}_{#1#2#3}}

\newcommand{\hhhhh}[5]{\mathrm{h}_{#1#2#3#4#5}}
\newcommand{\M}{N}

\begin{document}

\title{Analytic continuation of harmonic sums:\\[2mm] dispersion representation}

\author{\sc V.N.~Velizhanin\\[5mm]
\it Theoretical Physics Division\\
\it NRC ``Kurchatov Institute''\\
\it Petersburg Nuclear Physics Institute\\
\it Orlova Roscha, Gatchina\\
\it 188300 St.~Petersburg, Russia\\[2mm]
\it velizh@thd.pnpi.spb.ru
}

\maketitle

\begin{abstract}
We present a simple representation for analytically continued nested harmonic sums for the arbitrary complex arguments. This representation can be obtained for a wide range of nested harmonic sums from a precomputed database for the pole expressions of these sums near negative integers. We describe the procedure for the precise numerical evaluations of the corresponding results from the dispersion representation.

\end{abstract}

\section{Introduction}
\label{sec:intro}

The nested harmonic sums appeared from the first time during the calculations of the anomalous dimension of twist-2 operators in the second order of perturbative theory~\cite{Floratos:1977au,GonzalezArroyo:1979df} in the framework of Deep-Inelastic Scattering~(DIS). In fact, nested harmonic sums are a generalization of the well-known harmonic sum, i.e. partial sum of the harmonic series, which appear in the results for the leading order calculations~\cite{Gross:1973ju,Gross:1974cs}. 
When it was suggested, after some calculations in the third order~\cite{Larin:1996wd}, that a new type of nested harmonic sums would similarly appear in higher orders, the general aspects of nested harmonic sums were considered in Ref.~\cite{Vermaseren:1998uu,Blumlein:1998if}, including the general definition, algebraic and structural relations, synchronization, Mellin and inverse Mellin transforms, values at infinity and other. All this was implemented in the form of \texttt{summer}-package~\cite{Vermaseren:1998uu} for \texttt{FORM}~\cite{Vermaseren:2000nd,Ruijl:2017dtg}. 
The simplest harmonic sum and its generalization to nested harmonic sums can be written in the following recurrent way~\cite{Vermaseren:1998uu,Blumlein:1998if}
\begin{equation}
S_{a_1}(\M)
=\sum_{i_1=1}^\M \frac{({\rm sign}(a_1))^{i_1}}{i_1^{|a_1|}},\qquad
S_{a_1,a_2,a_3,\ldots}(\M)
=\sum_{i_1=1}^\M \frac{({\rm sign}(a_1))^{i_1}}{i_1^{|a_1|}} S_{a_2,a_3,\ldots}(i_1),
\label{HSR}
\end{equation}
where $a_i$ can be positive or negative, but not equal to zero and $\M$ is an integer.
The computations of heavy flavor corrections to deep-inelastic structure functions~\cite{Ablinger:2010ty}
required consideration of more general nested sums, the so-called \texttt{SSum}~\cite{Moch:2001zr} and cyclotomic sums, which were studied in Refs.~\cite{Ablinger:2011te,Ablinger:2012ufz,Ablinger:2013cf,Ablinger:2014bra}.

In real calculations of the anomalous dimensions of twist-2 operators through operator product expansion, calculations can be performed only for even (or only for odd) values of the Lorentz spin of operators. In practical use, it is sometimes necessary to know the result for other values (odd/even or complex). This can be done with the help of the inverse Mellin transform from nested harmonic sums in $\M$-space to harmonic polylogarithms in~$x$-space and the subsequent Mellin transform
\begin{equation}
S_{\vec{a}}(\M)\xrightarrow{\mathrm{inv.Mel.}} F_{S_{\vec{a}}}(x)\qquad \to \qquad
S_{\vec{a}}(M)=\int_0^1 d x\ x^{M} F_{S_{\vec{a}}}(x)
\label{MellinTr}
\end{equation}
with desired value for $M$.
The first step can be performed with the \texttt{harmpol}-package~\cite{Remiddi:1999ew} for \texttt{FORM} and the second step considered in Refs.~\cite{Blumlein:1997vf,Blumlein:2000hw,Blumlein:2005jg}.

Alternatively, one can perform the analytic continuation of the harmonic sums themselves, by generalizing the following procedure for the simplest harmonic sum
\begin{eqnarray}
S_1(\M)&=&\sum_{i=1}^\M\frac{1}{i}=\sum_{i=1}^\infty\frac{1}{i}-\sum_{i=\M+1}^\infty\frac{1}{i}
=\sum_{i=1}^\infty\frac{1}{i}-\sum_{i=1}^\infty\frac{1}{i+\M}\nonumber\\
&=&\sum_{i=0}^\infty\frac{1}{i+1}-\sum_{i=0}^\infty\frac{1}{i+\M+1}
=\Psi(\M+1)-\Psi(1)\,,\label{S1}
\end{eqnarray}
where $\Psi(z)$ is a digamma function, defined as the logarithmic derivative of the gamma function
\begin{equation}
\Psi(z)=\frac{d}{dz}\ln\Big(\Gamma(z)\Big)=\frac{\Gamma'(z)}{\Gamma(z)}
\end{equation}
which has the single poles at $z=0,-1,-2,\ldots$. A similar analytic continuation can be 
performed for any nested harmonic sum~(\ref{HSR}). For example, for the first non-trivial nested harmonic sum, which appear in the real computations~\cite{Floratos:1977au,GonzalezArroyo:1979df}, we can write, following Ref.~\cite{GonzalezArroyo:1979df}
\begin{eqnarray}
S_{-2,1}(\M)&=&\sum_{i=1}^\M\frac{(-1)^i}{i^2}S_1(i)
=\sum_{i=1}^\infty\frac{(-1)^i}{i^2}S_1(i)-\sum_{i=\M+1}^\infty\frac{(-1)^i}{i^2}S_1(i)\nonumber\\
&=&\sum_{i=1}^\infty\frac{(-1)^i}{i^2}S_1(i)-\sum_{i=1}^\infty\frac{(-1)^{i+\M}}{(i+\M)^2}S_1(i+\M)\nonumber\\
&=&S_{-2,1}(\infty)+(-1)^\M\sum_{i=1}^\infty\frac{(-1)^{i}}{(i+\M)^2}\Big[\Psi(i+\M+1)-\Psi(1)\Big]
\label{Sm21}
\end{eqnarray}
and now it is defined for all $\M$ and has poles at $\M=-1,-2,-3,\ldots$. The generalization of such procedures is described in Ref.~\cite{Kotikov:2005gr}. Advantages of this approach compared to the representation in the $x$-space is that it allows one to easily perform expansions in some limiting cases, such as $\M\to\infty$, which makes it possible to write a rather effective code for the evaluation of the nested harmonic sums for any values with high precision~\cite{Albino:2009ci}.

Following another approach from Ref.~\cite{Blumlein:2009ta}, one can consider mappings the nested harmonic sums into their related Mellin transform functions and, using the relations between harmonic sums with a given weight, express them all in terms of several such basis functions that are analytic at $x=1$. Performing an asymptotic expansion of the basis functions for $N\to\infty$ it is possible to obtain the numerical result for any complex value of $N$.

In the presented paper we show, how to obtain an alternative representation for nested harmonic sums using our database\footnote{The database is available on \url{https://github.com/vitvel/ACHS}}\cite{Velizhanin:2020avm} for $\omega$-expansion near negative integers $\M=-r+\omega,\ r=-1,-2,-3,\ldots$ without any additional calculations. Moreover, we provide the application of this dispersion representation for the numerical evaluation of nested harmonic sums for any (complex) values.

\section{Dispersion representation}\label{Section:1}

To obtain alternative representations for the analytically continued harmonic sums we will use as example the same nested harmonic sum $S_{-2,1}$ and we start with a simple code for \texttt{FORM} using the \texttt{harmpol}-package~\cite{Remiddi:1999ew}. We perform the inverse Mellin transform to $x$-space and expand it to obtain harmonic polylogarithms in $x$ and $\ln(x)$
\begin{verbatim}
#-
#include harmpol.h
L [S(R(-2,1),N)] = + S(R(-2,1),N) ;
* Inverse Mellin
#call invmel(S,N,H,x)
* Expand a general H-function into an expansion in x and ln(x)
#call hexpa(H,S)
B ln_;
print +s;
.end
\end{verbatim}
The output of this code contains the following three lines
\begin{equation}
\begin{aligned}
& \verb!  [S(R(-2,1),N)] =!\\
& \verb!         - sign_(N)*pow(x,i)*S(R(2),i)*sum(i,1,inf)*[1-x]*[1+x]^-1!\\
& \verb!         + sign_(N)*[1+x]^-1*z2!\\
& \verb!         - 5/8*z3!
\end{aligned}\label{Sm21FORM}
\end{equation}
The factor \texttt{sign\_(N)} in the right hand side indicates that the reconstruction from the even values of $\M$ is different from the reconstruction from the odd values: if we start from even values, we should set \texttt{sign\_(N)} to $(+1)$, but if we start from the odd values, we should set \texttt{sign\_(N)} to $(-1)$. Another additional multiplier \texttt{[1-x]*[1+x]\textasciicircum-1} can be transformed into the usual form with
\begin{eqnarray}
\sum_{i=1}^\infty \frac{1-x}{1+x}x^iS_{a_1,a_2,a_3,\ldots}(i)&\equiv&\sum_{i=1}^\infty(-x)^i S_{-a_1,a_2,a_3,\ldots}(i)\,,\qquad\label{xtominusxS}\\
\sum_{i=1}^\infty \frac{1-x}{1+x}x^i&\equiv&-\sum_{i=1}^\infty(-x)^i\,.\label{xtominusx}
\end{eqnarray}
Substituting Eq.~(\ref{xtominusxS}) into Eq.~(\ref{Sm21FORM}) and performing the Mellin transform
\begin{equation}
S_{-2,1}(\M) =
\mp \int_0^1 d x \sum_{i=1}^\infty (-x)^{i}x^\M S_{-2}(i)
\pm \int_0^1 d x\, x^{\M} \frac{\z2}{1+x}
- \frac{5}{8} \z3
\end{equation}
we obtain
\begin{equation}
S_{-2,1}(\M) =
\mp \sum_{i=1}^\infty \frac{(-1)^i}{i+N+1} S_{-2}(i)
\pm \z2 \sum_{i=1}^\infty \frac{(-1)^i}{i+N}
- \frac{5}{8} \z3\,,
\end{equation}
where $\pm$ signs correspond to the above \texttt{sign\_N}.
Using
\begin{equation}
\sum_{i=1}^\infty \frac{(-1)^i}{i+N+1} S_{-2}(i)=
\sum_{i=2}^\infty \frac{(-1)^{i+1}}{i+N} S_{-2}(i-1)=
\sum_{i=1}^\infty \frac{(-1)^{i+1}}{i+N} S_{-2}(i-1)
\end{equation}
as $S_{\vec{a}}(0)=0$, we will finally get
\begin{equation}
S_{-2,1}(\M) =
\pm \sum_{i=1}^\infty \frac{(-1)^i}{i+N} \Big[S_{-2}(i-1)
- \z2\Big]
- \frac{5}{8} \z3\,. \label{Sm21DR}
\end{equation}
Now this expression is correct for all complex values of $\M$, except $\M=-1,-2,-3,\ldots$.
We will call such representation, following L.~Lipatov\footnote{See \cite{DeVega:2001pu}; similar representation for the analytically continued harmonic sums was used in Refs.~\cite{Fadin:1998py,Kotikov:2000pm,Velizhanin:2015xsa,Velizhanin:2022seo} and in unpublished work of L.~Lipatov and A.~Onishchenko (2004).}, the dispersion representation. In the same way the dispersion representations can be obtained for other nested harmonic sums. However, below we will show, how this representation can be obtained in a simple way from our database~\cite{Velizhanin:2020avm} for the expansion of the analytically continued harmonic sums near negative integers, where they have poles.

In fact, the above dispersion representation is related to the expression obtained with the help of the usual analytic continuation of the nested harmonic sums~\cite{Gonzalez-Arroyo:1979guc,Yndurain:2006amm,Kotikov:2005gr,Ablinger:2013cf}, 
which in the case of the considered harmonic sum $S_{-2,1}$ looks like
\begin{eqnarray}
S_{-2,1}(\M)&\to&
\sum_{j=1}^{\infty}
\sum_{k=1}^{\infty}\Bigg[
\frac{\pm(-1)^k}{(k+\M)^2}\frac{1}{j+k+\M}
-\frac{\pm(-1)^k}{(k+\M)^2}\frac{1}{j}
-\frac{(-1)^k}{k^2}\frac{1}{j+k}
+\frac{(-1)^k}{k^2}\frac{1}{j}\Bigg]\qquad\label{UsualACSm21}
\\&=&
\sum_{j=1}^{\infty}
\sum_{k=1}^{\infty}
\frac{\pm(-1)^k}{(k+\M)^2}\frac{1}{j+k+\M}
-S_1^{\infty}\sum_{k=1}^{\infty}\frac{\pm(-1)^k}{(k+\M)^2}
-\zeta_{1,-2}
-\frac{\z2}{2} S_1^{\infty}\label{UsualACSm21S},
\end{eqnarray}
where $\zeta_{1,-2}=-\z2/2\,S_1^\infty+5/8\,\z3$.
Using the decomposition of the first term in the form
\begin{equation}
\frac{1}{(k+\M)^2}\frac{1}{j+k+\M}=-\frac{1}{j^2 (k+\M)}+\frac{1}{j^2 (j+k+\M)}+\frac{1}{j (k+\M)^2}
\end{equation}
it can be written as
\begin{equation}\label{rewritefirst}
\sum_{j=1}^{\infty}
\sum_{k=1}^{\infty}
\frac{1}{j^2}\frac{(-1)^k}{j+k+\M}
-\z2\sum_{k=1}^{\infty}\frac{(-1)^k}{(k+\M)}
+S_1^{\infty}\sum_{k=1}^{\infty}\frac{(-1)^k}{(k+\M)^2}
\end{equation}
Then, we change $j+k\to j$ for the first term and obtain 
\begin{equation}\label{resumfirst}
\sum_{j=1}^{\infty}
\sum_{k=1}^{\infty}
\frac{1}{j^2}\frac{(-1)^k}{j+k+\M}=
\sum_{j=2}^{\infty}
\frac{1}{j+\M}
\sum_{k=1}^{j-1}
\frac{(-1)^k}{(j-k)^2}\,,
\end{equation}
where
\begin{equation}\label{resummin}
\sum_{k=1}^{j-1}
\frac{(-1)^k}{(j-k)^2}=
(-1)^jS_{-2}(j-1)\,.
\end{equation}
Substituting Eqs.~(\ref{resummin}), (\ref{resumfirst}) and (\ref{rewritefirst}) into Eq.~(\ref{UsualACSm21S}) we obtain
\begin{equation}
S_{-2,1}(\M)\to\pm
\sum_{j=1}^{\infty}
\frac{(-1)^j}{j+\M}\Big(\HS_{-2}(j-1)-\z2\Big)
-\frac{5}{8}\z3\,,
\end{equation}
which is the same expression as Eq.~(\ref{Sm21DR}). 

\section{Dispersion representation from pole expressions}\label{Section:DisRep}

As we noted, the dispersion representation can be easily obtained from our database\footnote{The database is available on \url{https://github.com/vitvel/ACHS}}\cite{Velizhanin:2020avm}, which was computed by means of the inverse Mellin transform from the harmonic sums to the harmonic polylogarithms in $x$-space~with the help of the \texttt{harmpol}~\cite{Remiddi:1999ew} and \texttt{summer}~\cite{Vermaseren:1998uu} packages for \texttt{FORM}~\cite{Vermaseren:2000nd} and extraction of the small-$x$ logarithms, which correspond to poles in the initial $\M$-space.
Namely, we found the $\omega$-expansion of the analytically continued harmonic sums near negative integers $\M=-r+\omega,\ r=1,2,3,\ldots$, up to weight (or transcedentality) $13$ and up to $\zeta_{12}$ and similar, using available relations between multiple zeta-values from Ref.~\cite{Blumlein:2009cf}. One can see from Eq.~(\ref{Sm21DR}) that the right hand-side of this equation has simple poles for all $\M=-r,\ r=1,2,3,\ldots$. Our database~\cite{Velizhanin:2020avm} provides similar results for pole expressions of analytically continued harmonic sums.
For example, the simplest non-trivial nested harmonic sum with alternating summation has the following pole structure, being analytically continued from even integer $\M$ near $\M=-r,\ r=1,2,3,\ldots$ 
\begin{equation}\label{Sm21residues}
\check{S}_{-2,1}(r)=
\frac{(-1)^{r}}{\omega}\Big[S_{-2}(r-1)-\z2\Big]\,,\qquad \mathrm{vs.}\ \
\check{S}_{1}(r)=-\frac{1}{\omega}\,.
\end{equation}
Replacing $\omega\to j+M$ and $r\to j$ we obtain the dispersion representation for $S_{-2,1}(\M)$ as in Eq.~(\ref{Sm21DR})
\begin{equation}\label{Sm21T}
\hat{S}_{-2,1}(\M)=
\sum_{j=1}^{\infty}\frac{(-1)^{j}}{j+\M}\Big[S_{-2}(j-1)-\z2\Big]
+S_{-2,1}^{\infty},\qquad
S_{-2,1}^{\infty}=S_{-2,1}({\infty})=
-\frac{5}{8}\z3\,,
\end{equation}
where $\M$ is now an arbitrary complex number. The last term in Eq.~(\ref{Sm21T}) is fixed from the $\hat{S}_{-2,1}(0)=0$ condition, which gives
\begin{eqnarray}&&
-\sum_{j=1}^{\infty}\frac{(-1)^{j}}{j}\Big[S_{-2}(j-1)-\z2\Big]=
-\sum_{j=1}^{\infty}\frac{(-1)^{j}}{j}\Big[S_{-2}(j)-\frac{(-1)^j}{j^2}-\z2\Big]
\nonumber\\&&\qquad
=-S_{-1,-2}^\infty+S_{3}^\infty+S_{-1}^\infty\z2
=\z2\ln 2-\frac{13}{8}\z3+\z3-\z2\ln 2=-\frac{5}{8}\z3\equiv S_{-2,1}^\infty\,.\qquad
\end{eqnarray}
We observed, that this term must be equal to $S_{i_1,i_2,\ldots,i_n}^\infty$ for $\hat{S}_{i_1,i_2,\ldots,i_n}(\M)$ also in the general case.
For the analytic continuation from odd integers $\M$ we have to replace $(-1)^r$ to $-(-1)^r$ in Eq.~(\ref{Sm21residues}) and $(-1)^j$ to $-(-1)^j$ in Eq.~(\ref{Sm21}), while the last term must remain the same (with the same sign).

Thus, the dispersion representation for the analytic continuation of the nested harmonic sums can be performed from our database~\cite{Velizhanin:2020avm} for pole expressions with simple replacement rules, similar to the transformation from Eq.~(\ref{Sm21residues}) to Eq.~(\ref{Sm21T}). So, for the following nested harmonic sum with several alternating summation $S_{-2,-3,-2}$ we have from our database~\cite{Velizhanin:2020avm} the following pole expression for the analytic continuation of even values \footnote{Examples for the usage of our database~\cite{Velizhanin:2020avm} for such purposes are given in math-file of the \texttt{arXiv}-version this paper and on \url{https://github.com/vitvel/ACDR}.}
\begin{eqnarray}
\check{S}_{-2,-3,-2}(r)&=&
\frac{1}{2\, \w^3}\Big({\signN}-1\Big) {\z2} {\HS}_{-2}
+\frac{1}{8\, \w^2} \bigg(
{\signN} \Big(8{\HS}_{-2,-3}
-8 {\HS}_{5}\nonumber\\&&\qquad
+8 {\z2} {\HS}_{-3}
-12{\z3} {\HS}_{-2}
-3 {\z2} {\z3}\Big)
-8 {\z2} {\HS}_{-3}
+3 {\z2} {\z3}
\bigg)\nonumber\\&&
+\frac{1}{64\, \w} \bigg(
{\signN} \Big(128{\HS}_{-3,-3}
+192 {\HS}_{-2,-4}
+96 {\z2} {\HS}_{-4}
-192  {\z3} {\HS}_{-3}\nonumber\\&&\qquad
+168 {\z4} {\HS}_{-2}
-320 {\HS}_{6}
+180  \zeta_3^2
-303  {\z6}\Big)
-96 {\z2} {\HS}_{-4}
-147 {\z6}
\bigg),\quad
\end{eqnarray}
where $S_{\vec{a}}=S_{\vec{a}}(r-1)$. Then, we replace $\omega\to j+\M$ and $r\to j$ and obtain
\begin{eqnarray}
\hat{S}_{-2,-3,-2}(\M)&=&\sum_{j=1}^{\infty}\Bigg[
\frac{1}{2\, (j+\M)^3}\Big({\signj}-1\Big) {\z2} {\HS}_{-2}
+\frac{1}{8\, (j+\M)^2} \bigg(
{\signj} \Big(8{\HS}_{-2,-3}
\nonumber\\&&\qquad
-8 {\HS}_{5}
+8 {\z2} {\HS}_{-3}
-12{\z3} {\HS}_{-2}
-3 {\z2} {\z3}\Big)
-8 {\z2} {\HS}_{-3}
+3 {\z2} {\z3}
\bigg)\nonumber\\&&\hspace*{-20mm}
+\frac{1}{64\, (j+\M)} \bigg(
{\signj} \Big(128{\HS}_{-3,-3}
+192 {\HS}_{-2,-4}
+96 {\z2} {\HS}_{-4}
-192  {\z3} {\HS}_{-3}\nonumber\\&&\hspace*{-15mm}
+168 {\z4} {\HS}_{-2}
-320 {\HS}_{6}
+180  \zeta_3^2
-303  {\z6}\Big)
-96 {\z2} {\HS}_{-4}
-147 {\z6}
\bigg)\Bigg]+S_{-2,-3,-2}^\infty\,,\quad
\end{eqnarray}
where $S_{\vec{a}}=S_{\vec{a}}(j-1)$. For the analytic continuation of odd values of $\M$ we should replace only $(-1)^j$ to $-(-1)^j$.

\section{Numerical evaluation}

The dispersion representation provides the simple numerical evaluation of the analytically continued harmonic sums for arbitrary complex values. For the usual analytic continuation of nested harmonic sums one needs to do a lot of summations up to some upper limit of $N$ instead of infinity, that produce totally $N^w$ terms for $w$ summations. For the dispersion representation, $\M$-dependence is contained only in one summation, while other summations are collected into the usual nested harmonic sums. In this case the number of produced terms is suppressed by the factor $w!$, but is still large for the precision evaluation.
To speed up the computations of nested harmonic sums, one can use their expression through the corresponding linear difference equations\footnote{The linear difference equations for nested harmonic sums are used in \texttt{Sigma}-package~\cite{Schneider07symbolicsummation} to solve recurrences and reduce the initial sum to nested harmonic sums or similar.}. For these purposes we can use \texttt{FindSequenceFunction} from \texttt{MATHEMATICA}. For example, for one of the simplest non-trivial nested harmonic sum $S_{-2,1}$ we produced the sequence of values for at least the first 23 integers:
\begin{eqnarray}
\Bigg\{-1,\,
-\frac{5}{8},\,
-\frac{179}{216},\,
-\frac{1207}{1728},\,
-\frac{170603}{216000},\,
-\frac{155903}{216000},\,
-\frac{57395129}{74088000},\,
-\frac{433990957}{592704000},\,
\ldots
\Bigg\}
\end{eqnarray}
and applying \texttt{FindSequenceFunction} we get:
\begin{eqnarray}
&&\hspace*{-7mm}
\mathrm{Sm21DF}=\mathrm{DifferenceRoot}\Big [\,
\mathrm{Function}\Big [\,
\{\y,\n\},\nonumber\\&&\hspace*{-3mm}
\Big\{
-(\n+1)^2 (\n+2)\, \y[\n]
-(\n+2) (\n^2+7 \n+9)\, \y[\n+1]
+(\n^3+4 \n^2+\n-7)\, \y[\n+2]\nonumber\\&&\qquad
+(\n+3)^3\, \y[\n+3]
=0,\
\y[1]=-1,\
\y[2]=-\frac{5}{8},\
\y[3]=-\frac{179}{216}\
\Big\}\,
\Big]\,
\Big]\,.\label{Sm21DiffRoot}
\end{eqnarray}
The representation through the recurrence for the nested harmonic sum can be used to find its value for large integers in less time compared to its straightforward evaluation.
For example, the evaluation of $S_{-2,1}(\M)$ for $\M=10\,000$ in a usual way demands about $120$ sec., while the evaluation of~(\ref{Sm21DiffRoot}) demands about $50$ sec.. This gain in time is significantly increased by increasing the number of indices in a nested harmonic sum: the evaluation of $S_{-2,1,1,1,1,1}(\M)$, which enters into the four-loop anomalous dimension of twist-2 operators, for $\M=1\,000$ in the usual way demands about $30$ sec., while the evaluation of the recurrence relation demands less than $1$ sec.\footnote{Alternatively, one can use the package \texttt{HarmonicSums}~\cite{Ablinger:2012ufz,Blumlein:2009ta,Vermaseren:1998uu,Blumlein:1998if,Ablinger:2014bra,Ablinger:2014rba} for the exact evaluations of the harmonic sums, which demand even less time with compare to our method. However, our method allows to perform the numerical evaluation (see below), which is much faster.}.

To find the difference equations for all nested harmonic sums with weight $w$, we used our own code based on the assumption about the corresponding general ansatz in the following form:
\begin{equation}
S_{i_1,i_2,\ldots,i_m}(\M):\quad\sum_{i=0}^{w}a_i \,n^i+
\sum_{k=0}^{m}\sum_{i=0}^{w}a_{k,i}\, n^i\, y[n+k]\,,\qquad
y[\M]=S_{i_1,i_2,\ldots,i_m}(\M)\,.\label{AnsatzRE}
\end{equation}
Producing enough equations for various integers $\M$ we can solve the obtained system of linear equations for $a_{k,i}$. After substituting $a_{k,i}$ into Eq.~(\ref{AnsatzRE}) we obtain the linear difference equation in the form of Eq.~(\ref{Sm21DiffRoot}). In this way we performed similar calculations for all nested harmonic sums up to weight $7$ and for the nested harmonic sums without index $(-1)$ up to weight $11$\footnote{\label{note1}The math-file with corresponding results and examples can be found in the ancillary files of the \texttt{arXiv}-version this paper and on \url{https://github.com/vitvel/ACDR}.}.

For the evaluation of the analytically continued nested harmonic sums in the form of Eq.~(\ref{Sm21}) we need to precompute a lot of values for $S_{\vec{a}}(j-1)$. The most efficient way to do this inside \texttt{MATHEMATICA} is to use the function \texttt{RecurrenceTable}\footref{note1}. To generate $2\,000$ values for all $486$ nested harmonic sums with weight $6$, which is necessary for the evaluation of the harmonic sums with weight $7$, a typical laptop takes about $600$ sec. and about $2$ Gb of RAM. However, it is obvious that we do not need to know the exact results for $S_{\vec{a}}(j-1)$, since the final result will be numerical. Thus, we can evaluate all values numerically, which will speed up the calculations even more. For the numerical evaluation of the difference equations we must provide numerical values for the initial conditions (i.e., for example, the last three terms in Eq.~(\ref{Sm21DiffRoot})) and use \texttt{WorkingPrecision} option inside \texttt{RecurrenceTable}. 
To generate the same $2\,000$ values with \texttt{WorkingPrecision}$\,\to\! 100$ only 30 seconds and about $800$ Mb RAM  were needed, while for ten times more ($20\,000$) values we spent about $250$ sec. and about $4$ Gb RAM ($8$ Gb RAM for \texttt{WorkingPrecision}$\,\to\! 1000$).

The obtained database can be used for the numerical evaluation of the analytically continued harmonic sums within the dispersion representation similar to Eq.~(\ref{Sm21DR}) for $S_{-2,1}(\M)$. For example, we get from our database~\cite{Velizhanin:2020avm} the following dispersion representation for $S_{-2,1,1,1,1,1}$:
\begin{eqnarray}
S_{-2,1,1,1,1,1}(\M)&=&-
\sum_{j=1}^M\frac{\signN}{j+\M}\bigg[
S_{-5,1}
+S_{-4,2}
+S_{-3,3}
+S_{-2,4}
-S_{-4,1,1}
-S_{-3,1,2}
\nonumber\\&&\qquad
-S_{-3,2,1}
-S_{-2,1,3}
-S_{-2,2,2}
-S_{-2,3,1}
+S_{-3,1,1,1}
+S_{-2,1,1,2}
\nonumber\\&&\qquad
+S_{-2,1,2,1}
+S_{-2,2,1,1}
-S_{-2,1,1,1,1}
-S_{-6}
+\z6\bigg]+S_{-2,1,1,1,1,1}^\infty\,,\qquad\label{Sm211111DR}
\end{eqnarray}
where $S_{\vec{a}}=S_{\vec{a}}(j-1)$ and 
\begin{eqnarray}
S_{-2,1,1,1,1,1}^\infty&=&
\frac{7}{16} \hh31 \z3
-\frac{1}{24}\pi ^2 \hhh311
+\frac{1}{2}\hhhhh31111
+\frac{7}{34} \hhh331
-\frac{101}{68} \hhh511\nonumber\\&&
-\frac{1045}{39168} \pi ^4 \z3
-\frac{747}{4352} \pi ^2 \z5
+\frac{35767}{8704} \z7
\end{eqnarray}
with $\mathrm{h}_{i_1 i_2\ldots i_n}$ from Ref.~\cite{Blumlein:2009cf}, which can be calculated with the help of the relation
\begin{equation}
\mathrm{h}_{i_1 i_2 \ldots i_n}=
(-1)^n \mathrm{Li}\Big(\big\{|i_1|,|i_2|,\ldots,|i_n|\big\},
\big\{\mathrm{sgn}(i_1),\mathrm{sgn}(i_2),\ldots,\mathrm{sgn}(i_n)\big\}\Big)
\end{equation}
using the \texttt{GiNaC} implementation of Ref.~\cite{Vollinga:2004sn}.
We evaluated Eq.~(\ref{Sm211111DR}) for $\M=0$ and the results is given in Table~\ref{Table:1}.
From the second line of Table~\ref{Table:1} one can see, that the summation up to $M=5\,000$ gives the rather good accuracy about $10^{-4}$.

\begin{table}[h]
\renewcommand{\arraystretch}{1.5}
\centering
\begin{tabular}{|c|c|c|c|c|c|}
\hline
\hspace*{10mm}$M\ (\times 10^3)$\hspace*{10mm}& \hspace*{12mm} & 1 &\ \ 50\ \ \ &\ 100\ \ &\ 200\ \ \\
\hline
$S_{-2,1,1,1,1,1}$ $(\times 10^{-5})$ 
& $\M=0$    &   471   &  10.1    &  5.1    &  2.5   \\
\hline
$S_{2,1,1,1,1,1}$ $(\times 10^{-2})$ 
&$\M=0$& 35.5 & 2.7 & 1.6 & 1.0\\
\hline
${S}_{2,1,1,1,1,1}$ $(\times 10^{-8})$
&$\M=2$ & $47\cdot 10^3$ & 45.9 & 14.1 & 4.3 \\
\hline
$S_{-2,1,1,1,1,1}$ $(\times 10^{-11})$ 
&$\M=2$     &   $97\cdot 10^3$   &  40.5    &  10.1    &  2.5   \\
\hline
$S_{1,1,2,1,1}$ $(\times 10^{-8})$
&$\M=2$     &  $52\cdot 10^3$    &  69.0    &  20.5    &  6.0   \\
\hline
\end{tabular}
\caption{The difference between the exact result and the numerical evaluation of the dispersion representation for some nested harmonic sums.}
\label{Table:1}
\end{table}

However, if we consider the evaluation of the non-alternating sums, for example $S_{2,1,1,1,1,1}$, with the following dispersion representation
\begin{eqnarray}
&&\hspace*{-10mm}S_{2,1,1,1,1,1}(\M)=
\sum_{i=1}\frac{1}{j+\M}\bigg[
S_{6}
-S_{5,1}
-S_{4,2}
-S_{3,3}
-S_{2,4}
+S_{4,1,1}
+S_{3,1,2}
+S_{3,2,1}
+S_{2,1,3}
\nonumber\\&&\hspace*{-5mm}
+S_{2,2,2}
+S_{2,3,1}
-S_{3,1,1,1}
-S_{2,1,1,2}
-S_{2,1,2,1}
-S_{2,2,1,1}
+S_{2,1,1,1,1}
-\z6\bigg]+S_{2,1,1,1,1,1}^\infty\,,\label{S211111DR}
\end{eqnarray}
the result is much worse, as can be seen from the third line of Table~\ref{Table:1}, due to the bad-convergence of $(j+\M)^{-1}$ summation. Recall, however, that the last term in Eq.~(\ref{S211111DR}) $S_{2,1,1,1,1,1}^\infty=6\,\z7$ is fixed from the condition $S_{2,1,1,1,1,1}(0)=0$. To satisfy this condition, we can rewrite Eq.~(\ref{S211111DR}) as follows:
\begin{eqnarray}
&&\hspace*{-10mm}S_{2,1,1,1,1,1}(\M)=
\sum_{i=1}\left(\frac{1}{j+\M}-\frac{1}{j}\right)\!\bigg[
S_{6}
-S_{5,1}
-S_{4,2}
-S_{3,3}
-S_{2,4}
+S_{4,1,1}
+S_{3,1,2}
+S_{3,2,1}
\nonumber\\&&
+S_{2,1,3}
+S_{2,2,2}
+S_{2,3,1}
-S_{3,1,1,1}
-S_{2,1,1,2}
-S_{2,1,2,1}
-S_{2,2,1,1}
+S_{2,1,1,1,1}
-\z6\bigg],\label{S211111DRS}
\end{eqnarray}
where the second term in round brackets gives $S_{2,1,1,1,1,1}^\infty$, if we take into account, that $S_{k,\vec{a}}(i-1)=S_{k,\vec{a}}(i)-(-1)^iS_{\vec{a}}(i)/i^{|k|}$.
Now, $S_{2,1,1,1,1,1}(0)=0$ by definition and in the fourth and fifth lines of Table~\ref{Table:1} we present the difference for $S_{2,1,1,1,1,1}$ and $S_{-2,1,1,1,1,1}$ at $\M=2$ with their exact values and it can be seen that the accuracy has increased by six orders.

In the case, when the nested harmonic sum is started with index 1, for example $S_{1,1,2,1,1}$, we must first extract $S_1^n$ explicitly, which in the case of $S_{1,1,2,1,1}$ looks like 
\begin{eqnarray}
\HS_{1,1,2,1,1}&=&
\frac{1}{2} {\HS}_{1}^2 {\HS}_{2,1,1}
+{\HS}_{1} \Big(
{\HS}_{2,1,2}
+{\HS}_{2,2,1}
+{\HS}_{3,1,1}
-3 {\HS}_{2,1,1,1}
\Big)
+\frac{1}{2} {\HS}_{2,1,3}
+{\HS}_{2,2,2}
+\frac{1}{2} {\HS}_{2,3,1}\nonumber\\&&\hspace*{-15mm}
+{\HS}_{3,1,2}
+{\HS}_{3,2,1}
+\frac{1}{2} {\HS}_{4,1,1}
-\frac{5}{2} {\HS}_{2,1,1,2}
-\frac{5}{2} {\HS}_{2,1,2,1}
-2 {\HS}_{2,2,1,1}
-3 {\HS}_{3,1,1,1}
+6 {\HS}_{2,1,1,1,1}
   \label{Decomp}
\end{eqnarray}
and can be performed for arbitrary nested harmonic sum with the \texttt{HarmonicSums}-package~\cite{Ablinger:2012ufz,Blumlein:2009ta,Vermaseren:1998uu,Blumlein:1998if,Ablinger:2014bra,Ablinger:2014rba}.
After that, we replace $S_1$ with the polygamma function $\Psi(n)$
\begin{equation}
S_1(\M)\to \Psi(\M+1)-\Psi(1)\,,
\end{equation}
while other harmonic sums can be evaluated with the above procedure. In the sixth line of Table~\ref{Table:1} we give the difference with the exact result for $S_{1,1,2,1,1}(2)$.

One of the applications for the presented method is the numerical evaluation of the second and third order of BFKL Pomeron eigenvalues~\cite{Gromov:2015vua} for the conformal spin $n=0$ and BFKL intercept function for arbitrary conformal spin~\cite{Alfimov:2018cms}. From general consideration both results should coincide when the value of argument is equal to $(-1/2)$. Using the above described dispersion representation for the nested harmonic sums, entered into the corresponding expressions, in a short time we found\footnote{We used Eqs.(4.15) and (4.16) from \texttt{arXiv}-version of Ref.\cite{Alfimov:2018cms} and Eqs.(4) and (22) from \texttt{arXiv}-version of Ref.\cite{Gromov:2015vua} multiplied by $2$ due to Eq.(3) of Ref.\cite{Gromov:2015vua}. For the last two equations we should add terms that survive after the replacement $S_{\vec{a}}=0$, i.e., for example, $(-6\z3+4\pi^2\ln2)\times 2$ for Eq.(4) from Ref.\cite{Gromov:2015vua}.} nice agreement with the accuracy about $10^{-8}$ for $M=200\,000$ terms in sums.

\section{Conclusion}

In this paper we presented the dispersion representation for the analytically continued nested harmonic sums. This representation can be obtained from the available precomputed database~\cite{Velizhanin:2020avm} for the $\omega$-expansion of the analytically continued nested harmonic sums near negative integers $\M=-r+\omega,\ r=1,2,3,\ldots$, where they have the poles, with simple substitution rules given in Section~\ref{Section:DisRep} (see Eqs.~(\ref{Sm21residues}) and (\ref{Sm21T}) and below).
The main advantage of this representation is that it can be easily obtained from the pole expressions for analytically continued nested harmonic sums. If such information becomes available, we can immediately obtain the corresponding dispersion representation for the nested harmonic sum.

The dispersion representation can be used for the numerical evaluation of the nested harmonic sums for any (complex) value with the high accuracy\footnote{The math-file with corresponding results and examples can be found in the ancillary files of the \texttt{arXiv}-version this paper and on \url{https://github.com/vitvel/ACDR}.}. For these purposes, the calculations of nested sums for integers included in the dispersion representation were carried out through linear difference equations, which significantly speeds up the calculations as a whole. The presented method for the numerical evaluation gives high accuracy, but can be further improved along with the method from~\cite{Albino:2009ci}.

One of the most interesting applications of the dispersion representation is its use to study the relations between DGLAP and BFKL equations in $\mathcal{N}=4$ SYM theory, which was started in Ref.~\cite{Kotikov:2002ab}. The dispersion representation for the nested harmonic sums provides exactly the information that is needed for such analysis. Such study can help to extended the generalized double-logarithmic equation~\cite{Velizhanin:2022seo}, known at this moment for the case $N=-2+\omega$, to other values of $N$.

%%%%%%%%%%%%%%%%%%%%%%%%%%%%%%%%%%%%%%%%%%%%%%%%%%%%%%%%%%%%%%%%%%

 \subsection*{Acknowledgments}

The research was supported by a grant from the Russian Science Foundation No. 22-22-00803, https://rscf.ru/en/project/22-22-00803/.

\end{document}